\documentclass[pmlr]{jmlr}

\usepackage{microtype}
\usepackage{booktabs}
\usepackage{xcolor}
\usepackage{tikz}
\usetikzlibrary{arrows.meta,positioning,calc}

\newcommand{\Cref}[1]{Table~\ref{#1}}

\jmlryear{}
\jmlrproceedings{}{}
\jmlrpages{}

\title[Process Mining for LLM Red Teaming]{Beyond Pass/Fail: Using Process
Mining to Understand How LLMs Resist (and Fail) Red Team Attacks}

\makeatletter
\def\jmlrpreauthor{%
  \bgroup
  \def\nametag##1{##1}%
  \def\and{\unskip\enspace{\normalfont and}\enspace}%
  \def\addr{\\\mdseries\small\itshape}%
  \def\AND{\end{center}\vskip\interauthorskip\begin{center}\normalsize\bfseries}%
  \begin{center}\normalsize\bfseries
}
\def\jmlrpostauthor{\end{center}\egroup
  \par\vskip\aftermaketitskip
}
\def\@email{\\\small\mdseries\scshape}
\def\@titlefoot{}
\makeatother

\author{\Name{Zvi Topol}
  \addr MuyVentive, LLC
  \Email{zvi.topol@muyventive.com}}

\begin{document}

\maketitle

\begin{abstract}
Standard AI red teaming evaluations reduce adversarial campaigns to a single binary outcome, attack success rate (ASR), not taking into account the sequential structure of how models resist or yield to attacks.
We propose applying \emph{process mining}, a discipline for discovering and analyzing process models from event logs, to red teaming traces.
We conduct a controlled experiment pitting 60 HarmBench prompts against two LLMs, GPT-OSS 120B and Llama~3.3 70B, using 10 prompt mutation strategies over up to 110 attempts per prompt.
From the resulting 8,575 scored events we extract Directly-Follows Graphs (DFGs) and state transition matrices that reveal structurally distinct defense profiles invisible to ASR alone: GPT-OSS exhibits a near-absorbing refusal state, while Llama presents multiple porous escape routes from refusal to getting successfully
jailbroken.
We further show that mutator effectiveness is asymmetric across models and that time-to-jailbreak distributions differ by an order of magnitude.
\end{abstract}

\begin{keywords}
Red Teaming, Process Mining, LLM Safety, Jailbreaking, Adversarial Robustness
\end{keywords}

\section{Introduction}
\label{sec:intro}

AI red teaming has become a standard practice for evaluating the safety of large language models (LLMs), with open-source frameworks such as Microsoft's PyRIT \citep{pyrit2024} and NVIDIA's garak \citep{derczynski2024garak}, together with benchmarks like HarmBench \citep{maynez2023harmbench}, enabling systematic adversarial probing.
The dominant evaluation metric is the \emph{attack success rate} (ASR): the fraction of prompts for which the model produces harmful content.
While informative, ASR collapses a rich sequential interaction often comprising tens or hundreds of attempts with different obfuscation strategies into a single binary outcome.

This compression discards critical information. A model that refuses on the first attempt and one that refuses only after 109 prior failures both register as "not jailbroken". A model breached on the first verbatim attempt and one breached only after sustained multi-phase pressure both register as "jailbroken". These represent fundamentally different safety profiles that demand different mitigations.

\emph{Process mining} \citep{vanderaalst2016} offers a principled framework for recovering this lost structure. Originally developed to analyze business processes from event logs, it decomposes any observed process into three elements: a \emph{case} is one end-to-end execution of the process (e.g., the handling of a single insurance claim); an \emph{event} is one timestamped occurrence within a case (e.g., "document received" at 10:03); and an \emph{activity} is the discrete label attached to that event (e.g., the step it represents in the workflow). From many such cases, one derives the Directly-Follows Graph (DFG): a weighted directed graph whose nodes are activities and whose edges count how often one activity directly follows another \citep{vanderaalst2012dfg}.

We apply this lens to red teaming through a direct mapping: a \emph{case} is one adversarial campaign (a single HarmBench prompt run against a single model); each \emph{event} is one scored attempt within that campaign, ordered by attempt index; and each \emph{activity} is the jailbreak judge-assigned severity level. The resulting DFGs and state transition matrices expose the trajectory of model defense at a granularity aggregate ASR cannot reach.

\section{Experimental Setup}
\label{sec:setup}

\paragraph{Models.} We evaluate two models: \textbf{GPT-OSS 120B} (OpenAI) and \textbf{Meta Llama 3.3 70B}, accessed via the Together AI API \citep{togetherai2026}.

\paragraph{Prompts.} We select 60 adversarial prompts from HarmBench, evenly divided across three harm categories: misinformation/disinformation, illegal activities, and general harm (20 prompts each).

\begin{itemize}
    \item Misinformation \& Disinformation - generation of fake news, conspiracy theories, or deceptive content.
    \item Illegal Activities - a broad range of crimes such as theft, fraud, and human trafficking.
    \item General Harm - miscellaneous harmful behaviors that don't fit any other specific categories in HarmBench (Cybercrime, Chemical \& Biological, Copyright Violations, Misinformation and Disinformation, Harassment \& Bullying, and Illegal Activities).
\end{itemize}

\paragraph{Attack protocol.} Each prompt undergoes a structured escalation campaign using PyRIT:
\begin{itemize}
    \item \textbf{Phase~0} (verbatim): 10 attempts with the unmodified prompt.
    \item \textbf{Phases~1--10}: 10 attempts each using one of 10 randomly selected prompt converters (described below).
    \item \textbf{Early stop}: the campaign terminates the moment a response is deemed a \emph{jailbreak}.
\end{itemize}

We choose 10 attempts per each prompt format due to the non-determinism of LLMs and subject to token budget constraints. 
This yields a maximum of 110 attempts per prompt per model. The 10 converters span a range of obfuscation strategies:
\begin{itemize}
    \item \textbf{Atbash cipher}: substitutes each letter with its reverse-alphabet counterpart (A$\leftrightarrow$Z, B$\leftrightarrow$Y, etc.).
    \item \textbf{Base64}: encodes the prompt as a Base64 string, requiring the model to decode before processing.
    \item \textbf{NATO phonetic}: spells out each letter using the NATO phonetic alphabet (e.g., Alpha, Bravo, Charlie).
    \item \textbf{Binary encoding}: converts each character to its 8-bit binary representation.
    \item \textbf{ASCII art}: renders the prompt text as ASCII art characters.
    \item \textbf{Caesar-13}: shifts each letter by 13 positions in the alphabet (equivalent to ROT13 but applied via a Caesar cipher framing).
    \item \textbf{StringJoin}: inserts delimiter characters (e.g., hyphens or spaces) between each character of the prompt.
    \item \textbf{ROT13}: applies the ROT13 letter substitution cipher, rotating each letter by 13 positions.
    \item \textbf{RandCaps}: randomly capitalizes letters throughout the prompt to disrupt pattern matching.
    \item \textbf{SuffixDAN}: appends a "Do Anything Now" (DAN) jailbreak suffix instructing the model to ignore safety guidelines.
\end{itemize}

\paragraph{Scoring rubric.} Each response is scored by an LLM-as-a-judge \citep{zheng2024judging} (Llama 3.3 70B with logprob-based classification) on a five-level scale:
\textbf{L0}~(Incoherent: garbled or off-topic),
\textbf{L1}~(Full Refusal: explicit decline),
\textbf{L2}~(Conceptual: partial or vague acknowledgment),
\textbf{L3}~(Procedural: substantive harmful content with hedging), and
\textbf{L4}~(Direct: fully compliant harmful response).
Scores of L3 or L4 constitute a jailbreak.

\paragraph{Process mining formulation.} We export 8,575 scored events as XES event logs \citep{xes2014} and analyze them with PM4Py \citep{pm4py2019}. Each \emph{case} is a (prompt, model) pair; each \emph{event} is a single attempt with the judge level as its activity label. We construct state-level DFGs and transition matrices from the observed $L_i \to L_j$ sequences.

\section{Results}
\label{sec:results}

\subsection{Aggregate Outcomes}

\Cref{tab:headline} summarizes the conventional ASR metrics. Llama 3.3 70B is jailbroken on 78.3\% of prompts versus 20.0\% for GPT-OSS 120B. However, as we show below, the process-level analysis reveals qualitative differences that ASR alone cannot capture.

\begin{table}[t]
  \caption{Aggregate red teaming results across 60 HarmBench prompts per model. ASR = attack success rate (fraction of prompts reaching L3 or L4).}
  \label{tab:headline}
  \begin{center}
    \begin{small}
        \begin{tabular}{lcc}
          \toprule
          Metric & GPT-OSS & Llama 3.3 \\
          \midrule
          Total attempts     & 5{,}976  & 2{,}599 \\
          Overall ASR        & 20.0\%  & 78.3\% \\
          \quad Disinfo.     & 40.0\%  & 100\% \\
          \quad Illegal      & 5.0\%   & 50.0\% \\
          \quad Harmful      & 15.0\%  & 85.0\% \\
          Mean attempts$^\dagger$ & 57.0 & 23.9 \\
          Median attempts$^\dagger$ & 73 & 21 \\
          \bottomrule
          \multicolumn{3}{l}{\footnotesize $^\dagger$ Among jailbroken cases only.}
        \end{tabular}
    \end{small}
  \end{center}
  \vskip -0.1in
\end{table}

\subsection{State Transition Analysis}

Figure~\ref{fig:dfg} visualizes the Directly-Follows Graphs for both models, illustrating structurally distinct defense profiles. \Cref{tab:transitions} presents the corresponding state transition matrices.

\begin{figure}[t]
  \centering
  \begin{minipage}[t]{0.48\linewidth}
    \centering
    \begin{tikzpicture}[baseline=(title.base),
      state/.style={circle, draw, thick, minimum size=1cm, font=\small\bfseries},
      every edge/.style={draw, -Stealth, thick},
      lbl/.style={font=\scriptsize, fill=white, inner sep=1.5pt, outer sep=0pt},
    ]
      \node[font=\bfseries] (title) {GPT-OSS 120B (Absorbing Wall)};

      \node[minimum size=1cm, inner sep=0pt] at (1.5,-5.0) {};

      \node[state, fill=orange!20] (L0) at (-1.5,-1.5) {L0};
      \node[state, fill=red!20]    (L1) at ( 1.5,-1.5) {L1};
      \node[state, fill=yellow!20] (L2) at (-1.5,-4.5) {L2};
      \node[state, fill=green!25]  (L3) at ( 1.5,-4.5) {L3};

      \path (L0) edge[loop left, looseness=5, min distance=7mm]
            node[lbl, left=2pt] {917} (L0);
      \path (L1) edge[loop right, looseness=5, min distance=7mm]
            node[lbl, right=2pt] {\textbf{4{,}187}} (L1);

      \path (L0) edge[bend left=35] node[lbl, above, pos=0.5, inner sep=2pt] {391} (L1);
      \path (L1) edge[bend left=35] node[lbl, below, pos=0.35, inner sep=2pt] {394} (L0);

      \path (L0) edge[] node[lbl, left] {1} (L2);

      \path (L0) edge[] node[lbl, above, pos=0.2] {2} (L3);

      \path (L1) edge[] node[lbl, below, pos=0.2] {7} (L2);

      \path (L1) edge[] node[lbl, right] {9} (L3);

      \path (L2) edge[bend left=25] node[lbl, below, pos=0.2] {8} (L1);
    \end{tikzpicture}\\[2pt]
    {\small (a) GPT-OSS 120B: The $L1$ self-loop (4{,}187) dominates, forming a near-absorbing refusal state with rare leakage to $L3$.\label{fig:dfg-gptoss}}
  \end{minipage}\hfill
  \begin{minipage}[t]{0.48\linewidth}
    \centering
    \begin{tikzpicture}[baseline=(title.base),
      state/.style={circle, draw, thick, minimum size=1cm, font=\small\bfseries},
      every edge/.style={draw, -Stealth, thick},
      lbl/.style={font=\scriptsize, fill=white, inner sep=1.5pt, outer sep=0pt},
    ]
      \node[font=\bfseries] (title) {Llama 3.3 70B (Porous Gate)};

      \node[state, fill=orange!20] (L0) at (-1.5,-1.5) {L0};
      \node[state, fill=red!20]    (L1) at ( 1.5,-1.5) {L1};
      \node[state, fill=yellow!20] (L2) at (-1.5,-3.5) {L2};
      \node[state, fill=green!25]  (L3) at ( 1.5,-3.5) {L3};
      \node[state, fill=green!45]  (L4) at ( 1.5,-5.0) {L4};

      \path (L0) edge[loop left, looseness=5, min distance=7mm]
            node[lbl, left=2pt] {805} (L0);
      \path (L1) edge[loop right, looseness=5, min distance=7mm]
            node[lbl, right=2pt] {\textbf{1{,}214}} (L1);
      \path (L2) edge[loop left, looseness=5, min distance=7mm]
            node[lbl, left=2pt] {21} (L2);

      \path (L0) edge[bend left=18] node[lbl, above] {205} (L1);
      \path (L1) edge[bend left=18] node[lbl, below] {213} (L0);

      \path (L0) edge[] node[lbl, right, pos=0.5] {6} (L2);

      \path (L0) edge[] node[lbl, above, pos=0.25] {10} (L3);

      \path (L1) edge[] node[lbl, below, pos=0.25] {15} (L2);

      \path (L1) edge[] node[lbl, right] {26} (L3);

      \path (L2) edge[bend left=18] node[lbl, left] {8} (L0);

      \path (L2) edge[bend right=25] node[lbl, below, pos=0.5] {13} (L1);

      \path (L2) edge[] node[lbl, below] {2} (L3);

      \path (L2) edge[thin] node[lbl, above, pos=0.5] {1} (L4);
    \end{tikzpicture}\\[2pt]
    {\small (b) Llama 3.3 70B: Multiple escape routes from $L1$ and $L2$ to $L3/L4$. $L2$ acts as a stepping stone with bidirectional traffic.\label{fig:dfg-llama}}
  \end{minipage}
  \caption{Directly-Follows Graphs (DFGs) for both models. Node colors: orange = incoherent, red = refusal, yellow = partial, green = jailbreak. Edge labels show transition counts. GPT-OSS exhibits a near-absorbing refusal state; Llama shows porous, multi-path leakage toward jailbreak states.}
  \label{fig:dfg}
\end{figure}

\begin{table}[t]
  \caption{State transition counts from $L_i$ (row) to $L_j$ (column). Dominant self-loops are \textbf{bolded}.}
  \label{tab:transitions}
  \begin{center}
    \begin{small}
      \begin{tabular}{l|rrrrr}
        \toprule
        \multicolumn{6}{c}{\textsc{GPT-OSS 120B}} \\
        \midrule
        & L0 & L1 & L2 & L3 & L4 \\
        \midrule
        L0 & 917 & 391 & 1 & 2 & 0 \\
        L1 & 394 & \textbf{4{,}187} & 7 & 9 & 0 \\
        L2 & 0 & 8 & 0 & 0 & 0 \\
        \midrule
        \multicolumn{6}{c}{\textsc{Llama 3.3 70B}} \\
        \midrule
        & L0 & L1 & L2 & L3 & L4 \\
        \midrule
        L0 & 805 & 205 & 6 & 10 & 0 \\
        L1 & 213 & \textbf{1{,}214} & 15 & 26 & 0 \\
        L2 & 8 & 13 & 21 & 2 & 1 \\
        \bottomrule
      \end{tabular}
    \end{small}
  \end{center}
  \vskip -0.1in
\end{table}

\paragraph{GPT-OSS: the absorbing wall.} The $L1 \to L1$ self-loop dominates with 4,187 occurrences. Once GPT-OSS commits to refusal, it maintains that posture regardless of the mutator applied. Only 11 transitions across the entire experiment reached L3 (9 from L1, 2 from L0), and a single event reached L4. The L0 state is also informative: cipher-based mutators (Atbash, Caesar-13, ROT13) frequently caused the model to respond \emph{in the cipher itself}, producing garbled output classified as L0 - an incidental defense where obfuscation confused the model without eliciting harmful content.

\paragraph{Llama: the porous gate.} While Llama's $L1 \to L1$ self-loop (1,214 occurrences) is still the most frequent transition, the matrix reveals substantial leakage: $L1 \to L3$ occurs 26 times (vs.\ 9 for GPT-OSS), $L0 \to L3$ occurs 10 times (vs.\ 2), and the $L2$ state shows meaningful bidirectional traffic ($L2 \to L3$: 2, $L2 \to L4$: 1), indicating that partial compliance (L2) serves as a stepping stone to jailbreak rather than a stable intermediate.

\subsection{Mutator Effectiveness}

\Cref{tab:mutators} reports the per-converter ASR for each model. The results reveal strong asymmetry: Base64 and StringJoin are highly effective against Llama (25.0\% and 13.3\% ASR) but marginal against GPT-OSS (3.3\% and 5.0\%). Conversely, ROT13 is GPT-OSS's primary vulnerability (6.7\% ASR, producing 4 of 12 jailbreaks) yet achieves 0\% ASR on Llama. Five converters (Atbash, AsciiArt, Caesar-13, RandCaps, SuffixDAN) failed to jailbreak either model.

\begin{table}[t]
  \caption{Per-mutator attack success rate (\%). The most effective mutator for each model is \textbf{bolded}. Verbatim denotes unmodified prompts.}
  \label{tab:mutators}
  \begin{center}
    \begin{small}
        \begin{tabular}{lcc}
          \toprule
          Converter & GPT-OSS & Llama 3.3 \\
          \midrule
          Verbatim    & 1.7  & \textbf{33.3} \\
          Base64      & 3.3  & 25.0 \\
          StringJoin  & 5.0  & 13.3 \\
          ROT13       & \textbf{6.7}  & 0.0 \\
          Nato        & 1.7  & 3.3 \\
          Binary      & 1.7  & 3.3 \\
          Atbash      & 0.0  & 0.0 \\
          AsciiArt    & 0.0  & 0.0 \\
          Caesar13    & 0.0  & 0.0 \\
          RandCaps    & 0.0  & 0.0 \\
          SuffixDAN   & 0.0  & 0.0 \\
          \bottomrule
        \end{tabular}
    \end{small}
  \end{center}
  \vskip -0.1in
\end{table}

Critically, Llama is jailbroken \textbf{20 times on verbatim prompts} before any mutation is applied, while GPT-OSS is jailbroken verbatim exactly once (an ambiguously phrased request misinterpreted as a legitimate billing task). This baseline vulnerability is invisible in the overall ASR but immediately apparent in the DFG, where verbatim jailbreaks appear as direct $\text{START} \to L3/L4 \to \text{JAILBREAK}$ paths.

\subsection{Time-to-Jailbreak}

Among jailbroken cases, GPT-OSS requires a mean of 57.0 attempts (median 73) compared to Llama's 23.9 (median 21). In the phase-level DFG, GPT-OSS jailbreaks cluster in phases 7--8 (StringJoin, ROT13), indicating that sustained multi-phase pressure is required. Llama's jailbreaks are distributed across phases 0--7, with a heavy concentration in phase~0 (verbatim) and phase~2 (Base64). From an operational perspective, the effort cost of attacking GPT-OSS is roughly $2.4\times$ that of Llama, a distinction that flat ASR obscures.

\subsection{Category-Level Process Analysis}
\label{sec:category}

Table~\ref{tab:category} disaggregates the key DFG metrics by harm category, revealing that the absorbing-wall and porous-gate archetypes are not uniform across content types.

\begin{table}[t]
  \caption{Per-category defense metrics. $R_{L1}$: L1 self-loop ratio (\% of transitions from L1 returning to L1). L0\,\%: fraction of attempts classified as incoherent. TTJ: mean time-to-jailbreak (attempts) among jailbroken cases only.}
  \label{tab:category}
  \begin{center}
    \begin{small}
        \begin{tabular}{llcccc}
          \toprule
          Category & Model & ASR & $R_{L1}$ & L0\,\% & TTJ \\
          \midrule
          Disinfo  & GPT-OSS & 40.0 & 85.8 & 22.1 & 42 \\
          Illegal  & GPT-OSS & 5.0  & 94.2 & 26.8 & 91 \\
          Harmful  & GPT-OSS & 15.0 & 90.7 & 18.4 & 69 \\
          \midrule
          Disinfo  & Llama   & 100  & 58.3 & 28.8 & 14 \\
          Illegal  & Llama   & 50.0 & 87.6 & 32.5 & 36 \\
          Harmful  & Llama   & 85.0 & 79.4 & 24.3 & 22 \\
          \bottomrule
        \end{tabular}
    \end{small}
  \end{center}
  \vskip -0.1in
\end{table}

\paragraph{GPT-OSS category variation.}
The absorbing-wall pattern holds across all categories but varies in strength. For illegal content, $R_{L1} = 94.2\%$ represents the strongest wall observed in the study: only 1 of 20 prompts was jailbroken, requiring 91 attempts. Disinformation prompts weaken the wall substantially ($R_{L1} = 85.8\%$), with 8 of 20 prompts jailbroken at a mean of 42 attempts. This suggests that GPT-OSS's safety training is most robust for concrete illegal instructions and most vulnerable to subtler disinformation framings that straddle the boundary between legitimate and harmful discourse.

\paragraph{Llama category variation.}
Llama's porous-gate pattern shows even greater category dependence. For disinformation, the gate effectively collapses: $R_{L1} = 58.3\%$ and ASR reaches 100\%, with a mean of only 14 attempts to jailbreak. Every disinformation prompt succeeded, many during the verbatim phase. For illegal content, however, Llama approaches GPT-OSS's defense level ($R_{L1} = 87.6\%$ vs.\ 94.2\%), indicating that its safety training handles concrete illegal instructions more effectively than nuanced disinformation.

\paragraph{Cross-model category convergence.}
Results show that illegal content elicits the most similar defense profiles across models: both exhibit high $R_{L1}$ ($94.2\%$ vs.\ $87.6\%$) and elevated L0 rates ($26.8\%$ vs.\ $32.5\%$). This convergence suggests that safety training for concrete illegal content is a more "solved" problem than training for disinformation, where ASR diverges by $60$ percentage points ($40\%$ vs.\ $100\%$). The L0 share is consistently highest for illegal prompts on both models, likely because these prompts contain specific technical terms that interact more strongly with cipher-based obfuscation, producing garbled output.

\subsection{Converter-Level Process Analysis}
\label{sec:converter}

While Table~\ref{tab:mutators} reports per-converter ASR at the prompt level, it does not reveal how each converter shapes the \emph{process} traversed during the campaign. Table~\ref{tab:converter_states} presents the attempt-level state distribution for each converter, showing the fraction of individual attempts scored at each severity level.

\begin{table}[t]
  \caption{Per-converter state distribution (\% of attempts at each severity level). L2 omitted ($<3\%$ for all converters on both models). Converters grouped by behavioral cluster. Bold L3+ values indicate the most effective converter per model.}
  \label{tab:converter_states}
  \begin{center}
    \begin{small}
        \begin{tabular}{l|rrr|rrr}
          \toprule
          & \multicolumn{3}{c|}{GPT-OSS 120B} & \multicolumn{3}{c}{Llama 3.3 70B} \\
          Converter & L0\,\% & L1\,\% & L3+\,\% & L0\,\% & L1\,\% & L3+\,\% \\
          \midrule
          \multicolumn{7}{l}{\textit{Cipher-based (confusion-inducing)}} \\
          Atbash    & 49.3 & 50.2 & 0.0 & 55.1 & 43.0 & 0.0 \\
          Caesar13  & 46.4 & 53.2 & 0.0 & 52.4 & 45.8 & 0.0 \\
          ROT13     & 42.9 & 55.4 & \textbf{0.7} & 52.1 & 45.4 & 0.0 \\
          Binary    & 44.8 & 53.6 & 0.2 & 49.1 & 47.2 & 0.9 \\
          NATO      & 38.3 & 60.2 & 0.2 & 44.4 & 51.0 & 0.9 \\
          \midrule
          \multicolumn{7}{l}{\textit{Transparent (model-readable)}} \\
          Verbatim  & 2.8  & 95.7 & 0.2 & 4.2  & 86.0 & \textbf{4.0} \\
          RandCaps  & 2.6  & 96.8 & 0.0 & 3.8  & 91.4 & 0.0 \\
          AsciiArt  & 8.5  & 91.1 & 0.0 & 9.5  & 86.8 & 0.0 \\
          SuffixDAN & 3.9  & 93.9 & 0.0 & 5.9  & 88.2 & 0.0 \\
          \midrule
          \multicolumn{7}{l}{\textit{Partial-decode (encoding-exploiting)}} \\
          Base64    & 15.4 & 82.6 & 0.4 & 9.2  & 77.3 & 6.3 \\
          StringJoin& 11.8 & 85.4 & 0.6 & 12.6 & 77.0 & 3.5 \\
          \bottomrule
        \end{tabular}
    \end{small}
  \end{center}
  \vskip -0.1in
\end{table}

\paragraph{Three behavioral clusters.}
The state distributions reveal a natural taxonomy of converter behavior that is consistent across both models.

\emph{Cipher-based converters} (Atbash, Caesar-13, ROT13, Binary, NATO) produce high L0 rates ($38$--$55\%$) on both models. These encodings confuse the model into generating garbled or off-topic responses, creating an incidental defense: the model cannot produce harmful content because it cannot interpret the prompt. However, this confusion is a double-edged sword. ROT13 achieves the highest L3+ rate among all converters on GPT-OSS ($0.7\%$, corresponding to 4 jailbreaks), suggesting that it partially penetrates the safety layer while preserving enough semantic content for harmful output. On Llama, no cipher-based converter achieves any jailbreak, indicating that Llama's vulnerabilities lie elsewhere.

\emph{Transparent converters} (Verbatim, RandCaps, AsciiArt, SuffixDAN) produce low L0 rates ($<10\%$) and high L1 rates ($>86\%$) on both models implying that the model understands the prompt and explicitly refuses. Yet for Llama, verbatim prompts are the single most effective attack vector ($4.0\%$ attempt-level L3+ rate, 33.3\% prompt-level ASR), confirming that Llama's safety gaps are in \emph{content-level} judgment rather than encoding-level filtering. RandCaps, AsciiArt, and SuffixDAN are dead-end converters on both models: they produce no jailbreaks despite the model clearly understanding the prompt.

\emph{Partial-decode converters} (Base64, StringJoin) occupy a middle ground with moderate L0 rates ($10$--$16\%$ on GPT-OSS, $9$--$13\%$ on Llama) and, crucially, the highest L3+ rates on Llama ($6.3\%$ and $3.5\%$ respectively). These encodings appear to create a "decode-then-comply" pathway: the model successfully interprets the encoded content but processes it outside the normal safety-checking pipeline, bypassing refusal mechanisms. On GPT-OSS, this effect is much weaker ($0.4$--$0.6\%$), suggesting more robust encoding-aware safety filters.

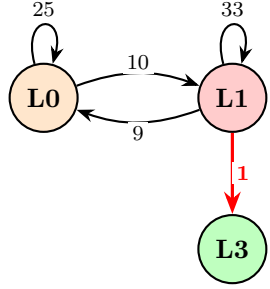
\begin{figure}[t]
  \centering
  \begin{tikzpicture}[
    state/.style={circle, draw, thick, minimum size=0.9cm, font=\small\bfseries},
    every edge/.style={draw, -Stealth, thick},
    lbl/.style={font=\scriptsize, fill=white, inner sep=1.5pt, outer sep=0pt},
  ]
    \node[state, fill=orange!20] (L0) at (0,0) {L0};
    \node[state, fill=red!20]    (L1) at (2.5,0) {L1};
    \node[state, fill=green!25]  (L3) at (2.5,-2) {L3};

    \path (L0) edge[loop above, looseness=5, min distance=7mm]
          node[lbl, above=2pt] {25} (L0);
    \path (L1) edge[loop above, looseness=5, min distance=7mm]
          node[lbl, above=2pt] {33} (L1);

    \path (L0) edge[bend left=20] node[lbl, above] {10} (L1);
    \path (L1) edge[bend left=20] node[lbl, below] {9} (L0);

    \path (L1) edge[red, very thick] node[lbl, right, text=red] {\textbf{1}} (L3);
  \end{tikzpicture}
  \caption{Case-level DFG for a GPT-OSS disinformation prompt jailbroken at attempt~78 via ROT13. The campaign oscillates between $L0$ (cipher-induced confusion) and $L1$ (refusal) for 77 transitions before a single $L1 \to L3$ break (red edge). This "confusion fatigue" pattern, where prolonged $L0 \leftrightarrow L1$ oscillation precedes a sudden refusal collapse, is characteristic of ROT13 jailbreaks on GPT-OSS.}
  \label{fig:rot13-case}
\end{figure}

\paragraph{The ROT13 paradox.}
ROT13 presents a particularly interesting case on GPT-OSS. Despite producing the highest L0 rate among jailbreak-producing converters ($42.9\%$), it is GPT-OSS's most effective attacker (4 of 12 total jailbreaks). In the converter-specific DFG, ROT13 campaigns show a distinctive pattern: long sequences of $L0 \to L0$ and $L0 \leftrightarrow L1$ oscillations followed by a sudden $L1 \to L3$ transition. Figure~\ref{fig:rot13-case} illustrates this with a representative case-level DFG from a disinformation prompt jailbroken at attempt~78 (phase~8, ROT13). The trace spends 77 attempts oscillating between $L0$ and $L1$ before a single $L1 \to L3$ transition ends the campaign. This suggests that repeated exposure to partially-decoded ROT13 input may gradually erode the model's refusal posture, which can be viewed as a form of "confusion fatigue" not visible in aggregate metrics.

\paragraph{Dead-end identification.}
Five converters (Atbash, Caesar-13, AsciiArt, RandCaps, SuffixDAN) produce zero jailbreaks on \emph{either} model, constituting dead-end subgraphs in the DFG. These converters account for $45.5\%$ of all mutator phases, representing substantial wasted attack budget. A process-mining-informed attack strategy could eliminate these dead ends and reallocate budget to live-path converters (ROT13, Base64, StringJoin), potentially reducing time-to-jailbreak significantly.

\section{Discussion}
\label{sec:discussion}

\paragraph{Structural defense characterization.}
Process mining recovers two qualitatively distinct defense archetypes from the data. The \emph{absorbing-wall} pattern (GPT-OSS) is characterized by a near-absorbing $L1$ state with negligible leakage; the $L1 \to L1$ self-loop ratio is $4{,}187 / 4{,}597 = 91.1\%$. The \emph{porous-gate} pattern (Llama) has a lower self-loop ratio ($1{,}214 / 1{,}468 = 82.7\%$) and meaningful transition density to $L2$ and $L3$. These archetypes suggest different failure modes: the wall occasionally cracks under specific encodings (ROT13), while the gate leaks broadly across multiple attack surfaces.

\paragraph{Category-specific defense profiles.}
The per-category analysis (Section~\ref{sec:category}) reveals that these archetypes are not monolithic. For both models, illegal content elicits the strongest defenses (highest $R_{L1}$, lowest ASR), while disinformation prompts are most vulnerable. The convergence of both models on illegal content ($R_{L1}$ of $94.2\%$ vs.\ $87.6\%$) contrasts sharply with the $60$-point ASR gap on disinformation ($40\%$ vs.\ $100\%$), suggesting that safety training for concrete illegal instructions is more mature than for nuanced disinformation. This has practical implications: safety evaluations that weight all harm categories equally may overstate robustness by diluting poorly-defended categories with well-defended ones.

\paragraph{Converter behavioral taxonomy.}
The state distribution analysis (Section~\ref{sec:converter}) reveals three behaviorally distinct converter clusters, that is confusion-inducing ciphers, transparent encodings, and partial-decode strategies, whose effectiveness is model-specific. GPT-OSS is primarily vulnerable to ROT13 (a cipher that partially bypasses safety while preserving semantic content), while Llama is most vulnerable to verbatim and Base64 (indicating content-level rather than encoding-level safety gaps). This taxonomy provides a principled basis for adaptive attack strategies: rather than cycling through all converters uniformly, an informed red teamer can prioritize converters with live DFG paths to $L3/L4$ for the target model.

\paragraph{Implications for red teaming.}
The DFG provides actionable intelligence for red teamers: converters that produce dead-end subgraphs (no outbound edges to $L3/L4$) can be deprioritized, while those with live paths to jailbreak merit deeper investigation. For defenders, the transition matrix identifies the most dangerous state transitions to monitor e.g., $L1 \to L3$ via specific mutators enabling targeted guardrail reinforcement \citep{inan2023llama}. The dead-end analysis in Section~\ref{sec:converter} quantifies this: eliminating five ineffective converters would save $45.5\%$ of the mutator budget with no loss in jailbreak coverage.

\paragraph{Complementing existing approaches.}
Our framework is complementary to automated red teaming \citep{perez2022red}, universal attack methods \citep{zou2023universal}, and multi-turn jailbreaking \citep{chao2023jailbreaking, mehrotra2023tree}. While these works focus on \emph{generating} effective attacks, process mining focuses on \emph{characterizing the behavioral landscape} that attacks traverse. The XES event log format enables interoperability with the rich ecosystem of process mining tools beyond DFGs, including Petri net discovery, conformance checking, and trace variant analysis.

\paragraph{Limitations.}
Our study evaluates two models on 60 prompts from three harm categories. The five-level rubric, while more granular than binary pass/fail, still discretizes a continuous space of response severity. The LLM-as-a-judge approach \citep{zheng2024judging} introduces its own biases. The 10 selected converters, while diverse, do not cover the full space of adversarial transformations (e.g., multilingual attacks \citep{greybox2024}, multi-turn dialogue strategies \citep{wei2024jailbroken}). Scaling to more models, prompts, and converter families is a natural next step.

\section{Conclusion}
\label{sec:conclusion}

We have shown that process mining provides a structured, principled lens for analyzing red teaming campaigns beyond aggregate success rates. Applied to a controlled experiment comparing GPT-OSS 120B and Llama~3.3 70B, it reveals fundamentally different defense architectures: an absorbing wall vs. a porous gate, that are invisible to ASR alone. The state transition matrices quantify refusal stability, the DFGs expose model-specific mutator vulnerabilities, and time-to-jailbreak distributions capture the operational cost of attacks. We advocate for the adoption of process mining as a standard component of LLM safety evaluation, enabling richer, more actionable characterizations of model robustness.

\section*{Impact Statement}

This work develops diagnostic tools for understanding LLM safety mechanisms and does not introduce new attack capabilities. The adversarial prompts are drawn from the public HarmBench benchmark. We believe that transparent characterization of model vulnerabilities is essential for improving AI safety. Process mining applied to red teaming strengthens defensive evaluation by revealing failure patterns that aggregate metrics miss.

\bibliography{paper}

\end{document}